\documentclass[11pt]{article}
\usepackage{latexsym}
\usepackage{amssymb}
\usepackage{amsfonts}
\usepackage{amsthm}
\usepackage{amsmath}
\usepackage{graphics, graphicx}     
\begin{document}


\begin{center}
{\Large \bf
2-D model of the global ionospheric conductor

connected with the magnetospheric conductors

\vspace{5mm}

Valery V. Denisenko$^{1,2}$
}
\end{center}

\vspace{2mm}

$^{1}$ {Institute of Computational Modelling, FRC "Krasnoyarsk Science Center"

Russian Academy of Sciences, Siberian Branch, 660036 Krasnoyarsk, Russia

denisen@icm.krasn.ru}

$^{2}$ {Siberian Federal University,
660041 Krasnoyarsk, Russia}

\vspace{4mm}

{\bf Abstract}

A model of the ionospheric global conductor is designed.
The ionospheric conductor is considered in the framework of a two-dimensional approximation based
on high conductivity in the direction of the magnetic field.
Under this assumption the magnetic field lines are equipotential,
and the charge transfer between them is determined only by integral Pedersen and Hall conductivities.
The model is constructed as the first approximation in the small parameter expansion of the solution
of the three-dimensional problems of electrical conductivity.
The small parameter is the ratio of Pedersen and field-aligned conductivities.
The space distributions of the Pedersen and Hall conductivities are calculated using the empirical models IRI, MSISE, IGRF
and applied to construct the maps of the integral conductivities.

The parts of the magnetosphere with high conductivity across the magnetic field lines,
namely, the cusps and the plasma layer are analyzed.
It is shown that the connection of these magnetospheric conductors
to the ionosphere in parallel makes the auroral zones equipotential regions.
As a consequence, for the ionospheric electric fields, which generators
are located in the ionosphere or in the atmosphere, the global problem of electrical conductivity is separated into three
independent boundary value problems in three regions: two polar caps and the main part of the ionosphere which includes
the mid- and low-latitude parts of the ionosphere.

The model can be used for the analysis of the ionospheric part of the Global Electric Circuit,
for calculation of the ionospheric dynamo electric field
and as a fragment in more complex ionospheric and magnetospheric models.

\vspace{3mm}

\section{Introduction}
\label{sec:Introduction}

For simulation of the quasistationary electric fields and currents the electrical conductivity equation is used.
It is set in section 2 for the three dimensional case.

The main input parameter for the 3-D electrical conductivity equation is the spatial distribution
of the conductivity, more precisely,
of the components of the conductivity tensor, since the geomagnetic field makes the conductivity of the ionospheric medium
equal to a gyrotropic tensor, whereas in the atmosphere the conductivity is a scalar.
The model of the conductivity of the ionosphere that we have created and its interface with the atmospheric conductivity is described
in section 3.
It is based on the empirical models of spatial distributions of the electron and
ion concentrations (IRI model), neutral molecules and atoms (MSIS model).
We calculate the geomagnetic field as the sum of $65$ spherical harmonics in accordance with the IGRF model.
We define the atmospheric conductivity as some combination of a few empirical models.

A 3-D model is usually reduced to a 2-D model
for mathematical simulation of the large-scale electric fields and currents in the ionosphere.
Two methods for such a simplifying the electrical conductivity problems are known.
Their comparison in \cite{SPb_2016_ion}
shows the advantage of the method based on a small parameter expansion. The small parameter equals to the ratio of the
conductivity across the magnetic field to the field-aligned conductivity.
Such a model was proposed in \cite{GurevichKrylovTsedilina}.
Our version is presented in \cite{Denisenko et al 2008b} with an emphasis on low latitudes.
In the simplest case when the geomagnetic field is vertical and the ionosphere is a homogeneous layer,
this small parameter expansion was verified in \cite{Justification}.

Within the framework of the 2-D model, obtained for infinite field-aligned conductivity, the magnetic field lines
are equipotential. The charges transfer along each magnetic field line freely, and
between the neighbor lines - due to the integral Pedersen and Hall conductivities.
In section 4 we construct the global distributions of the integral conductivities for the moment UT
$06$:$00$ in summer with minimal solar activity when Covington index $F10.7=80$.

Due to the high field-aligned conductivity the northern and southern parts of the ionosphere
are connected in parallel in the low and middle latitudes.
Since the magnetic field lines from the polar caps go into the lobes of the tail of the magnetosphere,
where the conductivity across the magnetic field is small,
we believe that the charges do not go above the polar caps.
The auroral zones require special consideration, since they are connected in parallel with the
plasma layer and cusps. This is done in section 5.

A 2-D equation is obtained from the 3-D charge conservation law in section 6.
The coefficients of this equation are the integral conductivities,
and its right-hand side is determined by external ionospheric currents and by currents,
flowing from the atmosphere.

In view of the equipotentiality of the auroral zones, established in section 5, the global electrical conductivity problem
splits into three independent boundary value problems which need to be solved in the polar caps and in the main part of the ionosphere,
which includes the mid- and low-latitude parts.
These three elliptic boundary-value problems are obtained in section 7.
Each of the problems has a unique solution. Our method of numerical solution is described in detail in \cite{Multigrid}.

\section{The electric conductivity equation} 
\label{sec:equations}

Here we regard the atmosphere, ionosphere and magnetosphere as a united conductor.

It is adequate to use a steady state model for a conductor with the conductivity tensor $\hat{\sigma}$
if the typical time of the process is much larger than the charge relaxation time
$\tau=\varepsilon_0/\sigma$ \cite{Molchanov et al 2008}.
Since atmospheric conductivity increases with height, it has minimal value near ground, where $\sigma > 10^{-14}\mbox{S/m}$
\cite{Molchanov et al 2008}.
So the charge relaxation time in the Earth's atmosphere is less than a quarter of an hour and
such a model can be used for atmospheric electric fields  which are not substantially varied during an hour or more.

The basic equations
for the steady state electric field
$\mathbf{E}$ and current density $\mathbf{j}$ are Faraday's law, the charge conservation law, and Ohm's law,
\begin{eqnarray}
\mbox{curl}\,\mathbf{E}=0,
\label{eq:1}\\
\mbox{div}\,\mathbf{j}=Q,
\label{eq:2}\\
\mathbf{j}=\hat{\sigma}\mathbf{E}.
\label{eq:3}
\end{eqnarray}

The equations (\ref{eq:1}, \ref{eq:2}) follow from Maxwell's four equations
when all parameters are time independent. The equation (\ref{eq:3}) is the empirical constitutive
equation between $\mathbf{j}$ and $\mathbf{E}$.
The given function $Q$ differs from zero if an external electric current exists.
Then the total current density is equal to $\mathbf{j}+\mathbf{j}_{ext}$ and the equation (\ref{eq:2})
with $Q=-\mbox{div}\,\mathbf{j}_{ext}$ is
the charge conservation law for the total current.

Because of the equation (\ref{eq:1}) the electric potential $V$ can be introduced so that
\begin{eqnarray}
\mathbf{E}=-\mbox{grad}\,V.
\label{eq:4}
\end{eqnarray}

Then the system of the equations (\ref{eq:1}-\ref{eq:3}) is reduced to the electric conductivity equation
\begin{eqnarray}
-\,\mbox{div}\left(\hat{\sigma}\,\mbox{grad}\,V\right)=Q.
\label{eq:5}
\end{eqnarray}

\section{Conductivity in the Earth's atmosphere and ionosphere} 
\label{sec:cond_ion_z}

We use parallel and normal to the direction of magnetic induction ${\bf B}$ components of vectors
which are marked with symbols $\parallel$ and $\bot$.
Then Ohm's law (\ref{eq:3}) in a gyrotropic medium takes the form
\begin{equation}
j_{_\parallel} = \sigma_{_\parallel}\*E_{_\parallel}, \qquad
{\bf j}_{_\bot}=\sigma_{_P}\*{\bf E}_{_\bot}-\sigma_{_H}\*\left[{\bf E}_{_\bot}\times{\bf B}\right]/ B,
\label{eq:1.2.2}
\end{equation}
with Hall ($\sigma_{_H}$) Pedersen ($\sigma_{_P}$) and field-aligned ($\sigma_{_\parallel}$)
conductivities \cite{Kelley}.

We have created the model \cite{Denisenko et al 2008b} to calculate the components
$\sigma_{_P}$, $\sigma_{_H}$, $\sigma_{_\parallel}$
of the conductivity tensor $\hat\sigma$ above $h=90$ km, that is based on the empirical models IRI, MSISE, IGRF.
In this model the ionospheric conductivity is calculated up to an altitude of 2000\,km. For our calculations we use the
profile up to the top of the ionospheric F--layer at $h_{_M}=500$\,km. If we include the layer above this height
the parameters of interest which are integral Pedersen and Hall conductivities would increase by only $1\,\%$.

Below $50$\,km the electric conductivity is isotropic. There exist many empirical models, for example
\cite{RycroftOdzimek}.
The conductivity does not depend of the magnetic field and so we can identify it as the field-aligned $\sigma_{_\parallel}$
conductivity.

At the heights $h=50-90$\,km the transformation from an atmospheric type of variation
to an ionospheric one occurs \cite{RycroftOdzimek, Schumann}.
We approximate a height dependance in the upper atmosphere as a smooth continuation
from the ionosphere above $h=90$ km to the values below $50$\,km which are typical for the atmosphere.
Namely, in the layer $50$\,km $<h<90$\,km,
the values for $\log{\sigma_{_\parallel}}$ and $\log{\sigma_{_P}}$ are interpolated by cubic functions of $h$.

\begin{figure}[h]
\hspace*{30mm}
\includegraphics[scale=1.0]{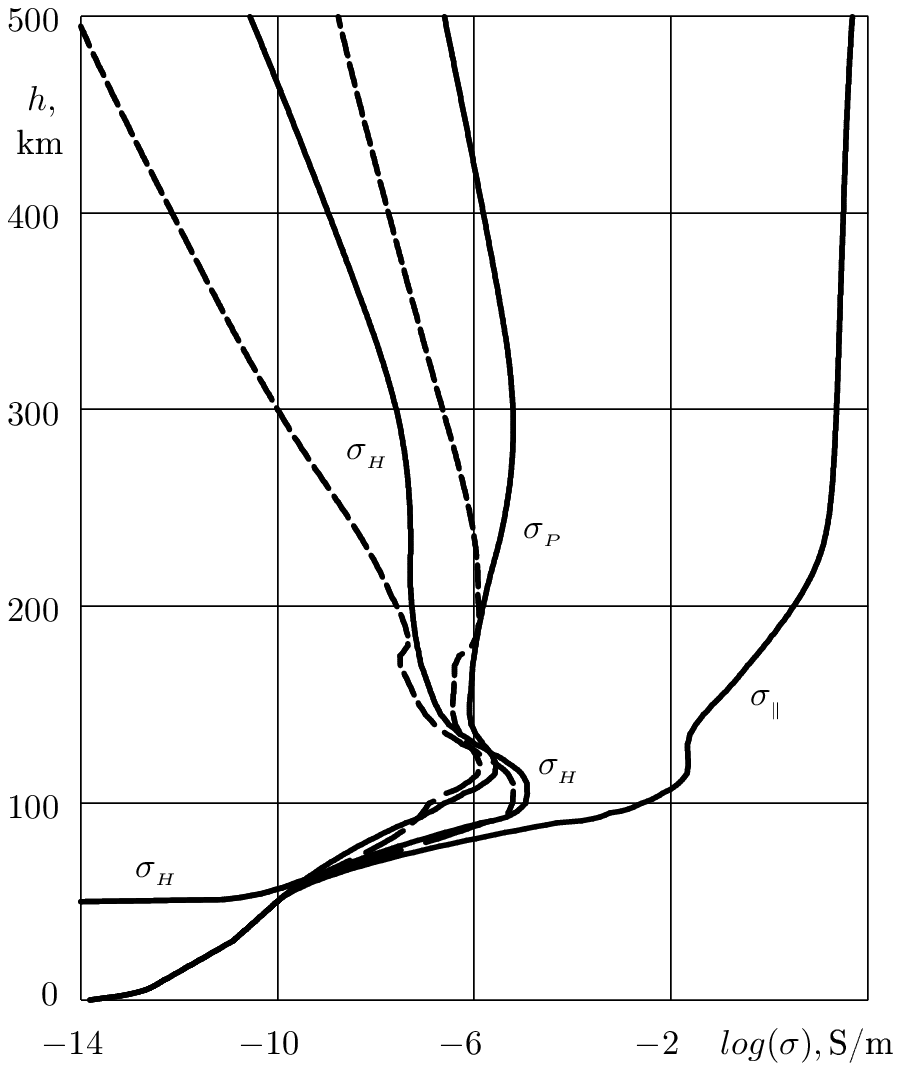}
\caption{
Profiles of the components of the electric conductivity tensor for a mid-latitude night--time ionosphere.
Plotted are the field-aligned conductivity $\sigma_{_\parallel}$,
the Pedersen conductivity $\sigma_{_P}$, and the Hall conductivity $\sigma_{_H}$ (solid lines).
The effective Pedersen and Hall conductivity averaged during acceleration period of $3$\,hours
are presented by the dashed lines.
}
\label{fig:sigma}
\end{figure}

The model \cite{Denisenko et al 2008b} permits us to calculate conductivities in the ionosphere
only above $h=80$\,km since the model IRI is not applicable below this height. These calculations show
that all components of the conductivity tensor are defined by electrons below $h=90$\,km and all ions give
negligible contributions.
We suppose that such a domination takes place also in the whole layer $50$\,km $<h<90$\,km where the values
for $\sigma_{_\parallel}$ and $\sigma_{_P}$ we obtain by continuation of the ionospheric height distributions.

For plasma with one dominating kind of charged particles the formulae for conductivities written
in \cite{Hargreaves} are simplified.
Then they give the following relation between components of the conductivity tensor
\begin{eqnarray}
\sigma_{_H}(h)=\sqrt{\sigma_{_P}(h)\left[\sigma_{_\parallel}(h)-\sigma_{_P}(h)\right]}.
\label{eq:Hall}
\end{eqnarray}

So it is not necessary to interpolate the values for $\sigma_{_H}$. It can be deduced from this formula after
interpolation of $\sigma_{_\parallel}$ and $\sigma_{_P}$.
The Hall parameter $\sigma_{_H}/\sigma_{_P}$ approximately equals to the ratio between
the electron gyrofrequency and the electron-neutral collision frequency.
As Fig. \ref{fig:sigma} shows, it takes a value of about $25$ at the height $90$ km.
For $\sigma_{_\parallel}>>\sigma_{_P}$ the formula (\ref{eq:Hall})
means $\sigma_{_\parallel}/\sigma_{_H}\simeq\sigma_{_H}/\sigma_{_P}$.
Since such an equality is valid at the height $85-95$ km we can use this approximation.
We extrapolate it down to $50$ km.
In our model the Hall parameter equals zero below $50$ km which corresponds to isotropic conductivity there.

The typical mid-latitude height distributions are shown in Fig. \ref{fig:sigma} for night--time conditions in
summer under minimal solar activity. It should be mentioned that these are averaged profiles
and the values of actual conductivity on a particular day can be a few times different.

The dashed lines in Fig. \ref{fig:sigma} present the effective Pedersen and Hall conductivities, which
describe the ionospheric conductor accelerated by Ampere's force. Such an acceleration would make the conductor
move with a drift velocity if the time is long enough and no other force exists.
Here we use an averaged acceleration period of $\tau_{_A}=3$\,hours.
A detailed explanation can be found in \cite{Denisenko et al 2008b}.
Sometimes this effect is taken into account in a simplified form as neglecting
$\sigma_{_P}, \sigma_{_H}$ above $160$ km \cite{Forbes}.
It is not adequate for $\sigma_{_P}$ in the night--time ionosphere as can be seen in Fig. \ref{fig:sigma}.

\section{2-D model of the ionospheric conductor}
\label{sec:2-D_ionosphere} 

It is shown in \cite{Hargreaves} how to reduce a three-dimensional model to a two-dimensional one when
the conductivity in the direction of the magnetic field $\sigma_{_\parallel}$ is
a few orders of magnitude larger than $\sigma_{_P}, \sigma_{_H}$.
We follow the approach of \cite{GurevichKrylovTsedilina} where this procedure is made
accurately from the mathematical point of view.
Our simplified version of this type of model is presented in \cite{Denisenko et al 2008b}.
A similar approach near geomagnetic equator is used in \cite{Kartalev et al 2006}.
Here we briefly present the key features of the model.

As can seen in Fig. \ref{fig:sigma} the conductivity in the direction
of the magnetic field $\sigma_{_\parallel}$ is a few orders of magnitude larger than $\sigma_{_P}, \sigma_{_H}$
in the layer where $\sigma_{_P}, \sigma_{_H}$ are large.
It is possible to idealize this inequality as
\begin {equation}
\sigma_{_\parallel}=\infty
\label{eq:14}
\end {equation}
in some layer $h_{_I}<h<h_{_M}$ for which parameters $h_{_I}, h_{_M}$ are to be chosen.

The equality (\ref{eq:14}) means that the electric current along a magnetic field line can be arbitrary,
while the electric field component $E_{_\parallel}$ equals zero,
\begin {equation}
E_{_\parallel}=0.
\label{eq:15}
\end {equation}

Because of (\ref{eq:4}, \ref{eq:15}) the electric potential $V$ is constant at each magnetic field line and
\begin {equation}
{\bf E}_{\perp}=-\mbox{grad}_{\perp} V.
\label{eq:16}
\end {equation}

\begin{figure}[h]
\hspace*{15mm}
\includegraphics[scale=0.8]{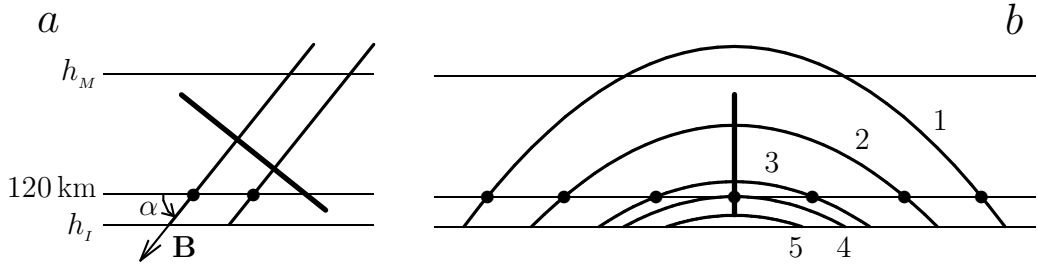}
\caption{
Magnetic field lines in the ionospheric layer between the heights $h_{_I}$ and $h_{_M}$
where conductivities $\sigma_{_P}$, $\sigma_{_H}$ are large.
a - for middle-latitudes, b - for equatorial ionosphere.
$\alpha$ - magnetic inclination.
Geomagnetic coordinates of the plotted black dots are used
to identify in global pictures the ionospheric parts of the lines which contain these dots.
Dark segments - cross-sections of possible coordinate surfaces which can be used in 2-D models.
}
\label{fig:lines_ionosp}
\end{figure}

Two such equipotential segments are shown in Fig. \ref{fig:lines_ionosp} a for middle-latitudes.
Panel b shows the equatorial ionosphere.
A couple of magnetic field lines separate the cross-sections of magnetic field tubes which are analyzed below.

Since each magnetic field line is an equipotential the ionospheric conductor
may be represented by Pedersen and Hall conductances which are equal to integrals of the corresponding local
conductivities $\sigma_{_P}, \sigma_{_H}$ \cite{Hargreaves}.

In such a model, a magnetic field line has its own value of the electric potential $V$.
It can obtain or lose charge by currents ${\bf j}_{_\perp}$ and it does not matter for its total charge
at what point along the magnetic
field line ${\bf j}_{_\perp}$ exists, because charges can go freely along the line according to
infinite $\sigma_{_\parallel}$ (\ref{eq:14}).

The electric field ${\bf E}_{_\perp}$ produces the current ${\bf j}_{_\perp}$; by Ohm's law (\ref{eq:1.2.2}) this is given by
\begin{equation}
{\bf j}_{_\perp}=
\left(\begin{array}{cc}
\sigma_{_P} & -\sigma_{_H} \\
\sigma_{_H} &  \sigma_{_P}
\end{array}\right)
{\bf E}_{_\perp}.
\label{eq:20}
\end{equation}

By summation of the inputs from all points of the magnetic field line, we obtain the conductance
between magnetic field lines. The resulting Ohm's law can be written as
\begin{equation}
{\bf J}_{_\perp}={\bf \hat\Sigma}\,{\bf E}_{_\perp},
\label{eq:20.1}
\end{equation}
where ${\bf J}_{_\perp}$ is the total current across magnetic field line.

If the magnetic field lines are parallel straight lines, then ${\bf E}_{_\perp}$ is constant in this integration and so
\begin {equation}
{\bf J}_{_\perp}=(\int
\left(\begin{array}{cc}
\sigma_{_P} & -\sigma_{_H} \\
\sigma_{_H} &  \sigma_{_P}
\end{array}\right)dl\,)\,{\bf E}_{_\perp},
\label{eq:27}
\end {equation}
which permits us to write down the tensor ${\bf \hat\Sigma}$ as
\begin {equation}
{\bf \hat\Sigma}=
\left(\begin{array}{cc}
\Sigma_{_P} & -\Sigma_{_H} \\
\Sigma_{_H} &  \Sigma_{_P}
\end{array}\right),
\label{eq:28}
\end {equation}
with Pedersen and Hall conductances $\Sigma_{_P}, \Sigma_{_H}$ which are obtained from the
local Pedersen and Hall conductivities $\sigma_{_P}, \sigma_{_H}$ by integration along a magnetic field line
\begin {equation}
\Sigma_{_P}=\int \sigma_{_P}\, dl, \qquad \Sigma_{_H}=\int \sigma_{_H}\, dl.
\label{eq:29}
\end {equation}

\begin{figure}[h]
\includegraphics[scale=0.9]{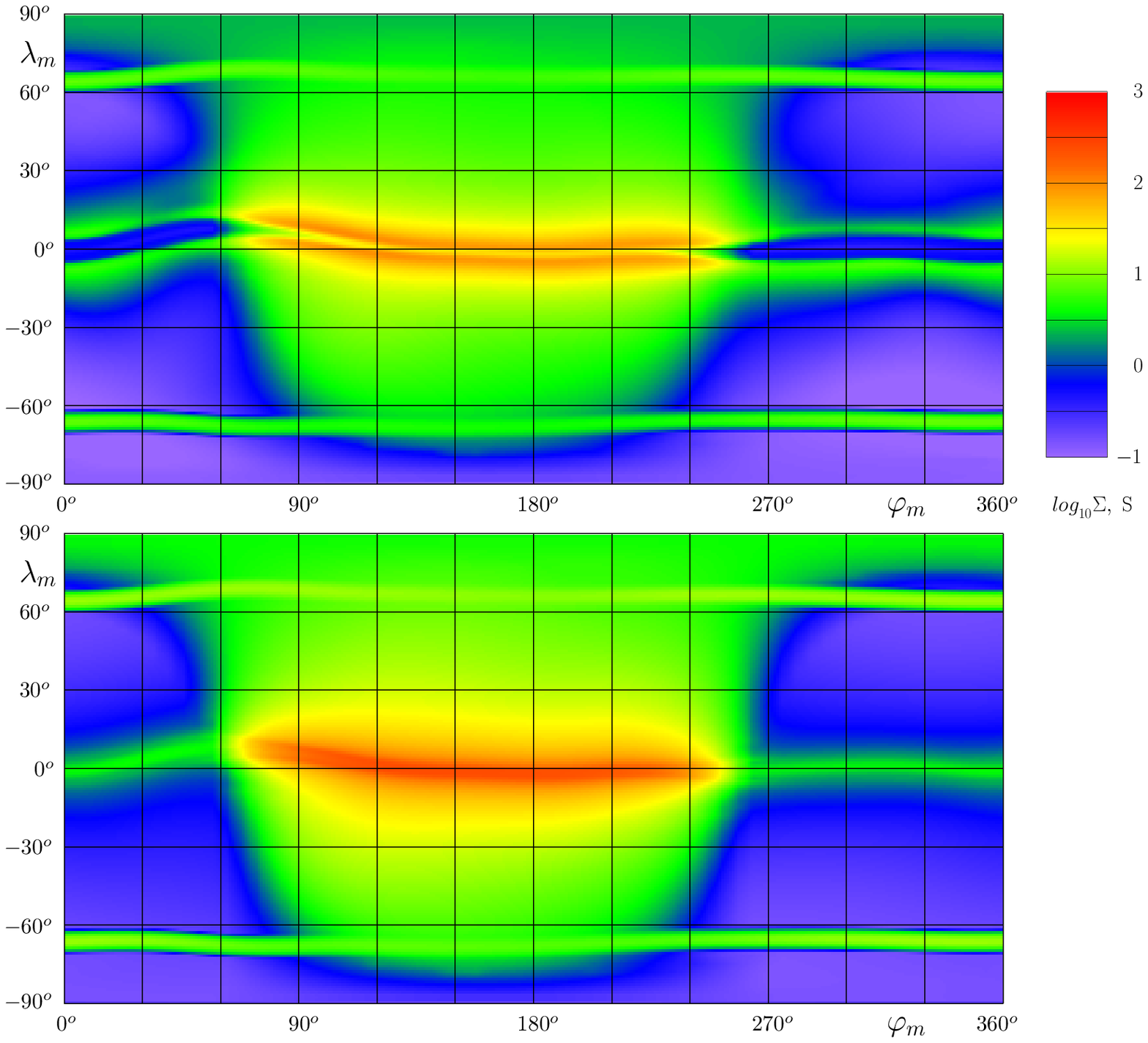}
\caption{
Distribution of the integral Pedersen conductance $\Sigma_{_P}$ (top panel) and Hall conductance
$\Sigma_{_H}$ (bottom panel).
The points with $\lambda_m, \varphi_m$ geomagnetic coordinates at 120 km height
in the ionosphere identify halves of magnetic field lines.
Maps are calculated under typical conditions for July under minimal solar activity
at the considered point in time, $06$:$00$ UT.
}
\label{fig:sig_120}
\end{figure}

Of course it is not necessary to integrate along the whole magnetic field line because of small $\sigma_{_P}, \sigma_{_H}$
outside some layer $h_{_I}<h<h_{_M}$.
As we already wrote $h_{_M}=500\,\mbox{km}$ can be taken as the upper boundary.
We use $h_{_I}=80\,\mbox{km}$ because of small conductivity below this height.
Calculations show that inclusion of conductivity outside this layer
would increase the integral Pedersen and Hall conductivities by less than $1\,\%$, which is negligible.

The empirical model IRI does not present any auroral enhancement of electron concentration
that is produced by high energy electron and proton precipitation from the magnetosphere.
Corresponding enhancement of conductivity is usually added as the auroral zones with
large integral conductances $\Sigma_{_P}, \Sigma_{_H}$.
These values are rather variable. We use some average values of the models
\cite{KamideMatsushita1, Spiro, Weimer}.
Namely we increase $\Sigma_{_P}, \Sigma_{_H}$
along the auroral oval whose central line has geomagnetic latitude $67^o$ at midday and $65^o$ at midnight
with values $\Sigma_{_P}=2$\,S and $8$\,S in these points. The half width of the ring equals $\delta\theta_m=5^o$.
Smooth interpolation is used to get the values in all points within auroral zone.
The same enhancement is done in the Southern hemisphere.
Additional Hall conductance in the auroral zone can be approximately taken as
\begin {equation}
\Sigma_{_H}^{aur}(\theta_m,\varphi_m)=1.5\,\Sigma_{_P}^{aur}(\theta_m,\varphi_m).
\label{eq:34}
\end {equation}
in accordance with \cite{Weimer}.

The obtained global distributions of $\Sigma_{_P}, \Sigma_{_H}$ are presented in Fig. \ref{fig:sig_120}.
Logarithmic scale is used since the values vary by almost four orders of magnitude.
The integral conductance $\Sigma_{_P}$ or $\Sigma_{_H}$ at each half of a magnetic field line
is shown in the dot where it crosses the surface $h=120\,\mbox{km}$.
In other words a half of a magnetic field line is substituted with a dot as
is shown in Fig. \ref{fig:lines_ionosp}.
It must be mentioned that equatorial magnetic field lines which are below $h=120\,\mbox{km}$
as the line 5 in Fig. \ref{fig:lines_ionosp} b, are absent in those pictures.
We would like to stress that it is a problem of visualization only and does not exist in calculations.

\begin{figure}[h]
\hspace*{5mm}
\includegraphics[scale=0.85]{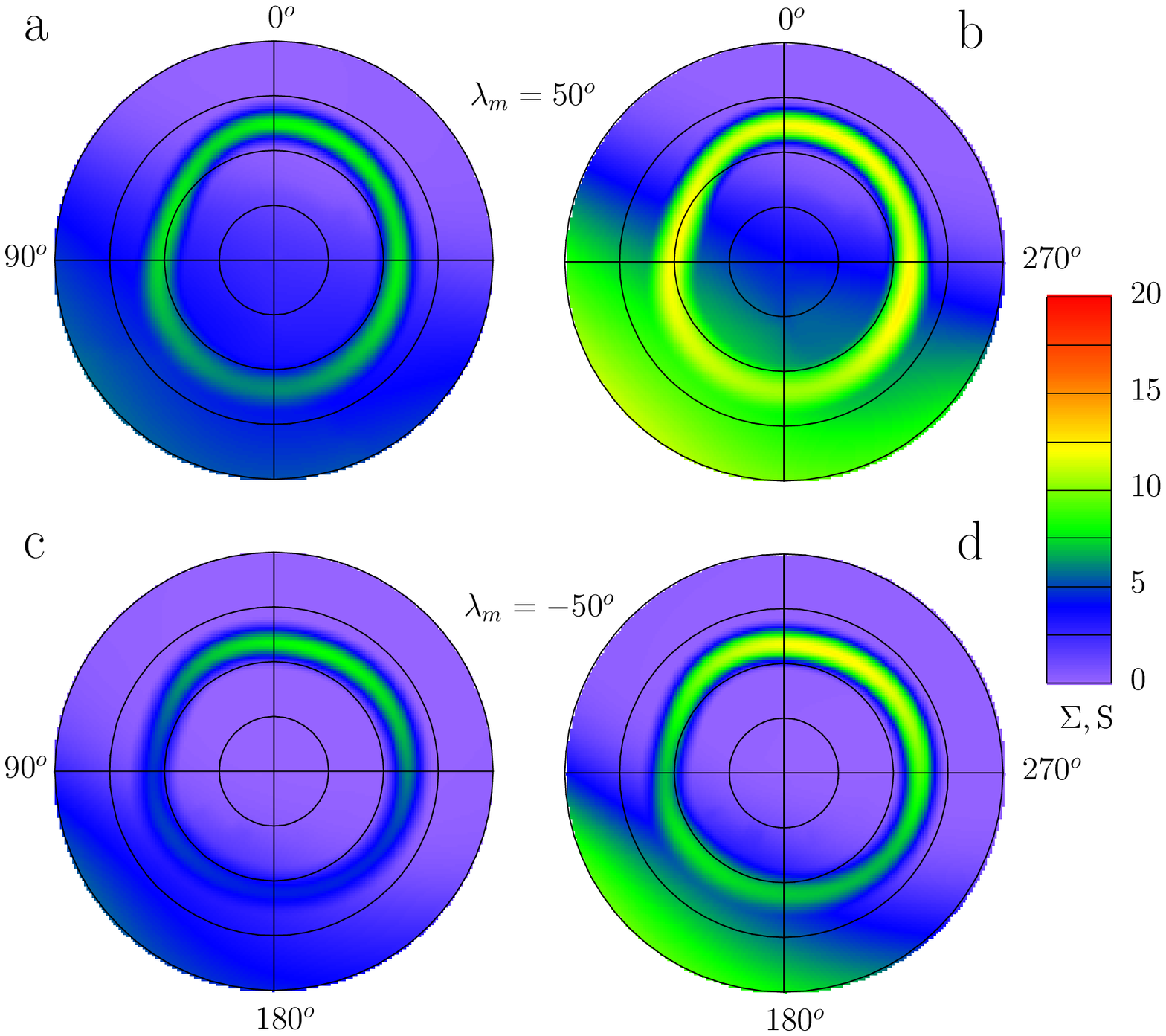}
\caption{
Distribution of the integral Pedersen $\Sigma_{_P}$ (a, c) and Hall $\Sigma_{_H}$ (b, d) conductances in the
Northern (a, b) and Southern (c, d) polar regions.
Maps are calculated under typical conditions for July under minimal solar activity
at the considered point in time, $06$:$00$ UT.
}
\label{fig:sig_120_caps}
\end{figure}

Fig. \ref{fig:sig_120_caps} shows high latitude fragments of the same $\Sigma_{_P}, \Sigma_{_H}$.
Both Northern and Southern fragments are shown as they look from the Northern pole.

At low geomagnetic latitudes a magnetic field line can be separated into
Northern and Southern halves only conventionally.
We use the apex of the line for such a separation during integration (\ref{eq:29}).
Each half line can be identified by geomagnetic coordinates $\lambda_m,\varphi_m$ of its point at some height
as is shown in Fig. \ref{fig:lines_ionosp}.
We use $120\,\mbox{km}$ for presentation in Fig. \ref{fig:sig_120}, \ref{fig:sig_120_caps}.
Of course the lines whose apexes are below this height like the line 5 in Fig. \ref{fig:lines_ionosp} b are not presented there.
If we choose a surface $100$\,{km} or
lower these short half lines with small values of conductance
would appear and produce a singularity in the picture.
Both halves of the line 4 in Fig. \ref{fig:lines_ionosp} have the same dot for their identification and the values of
$\Sigma_{_P}, \Sigma_{_H}$ at these halves are a little bit different, but this jump at the equator
is not seen in the pictures.

It is better to use a vertical surface for identification of the magnetic field lines near the geomagnetic equator.
Such a presentation of $\Sigma_{_P}, \Sigma_{_H}$ is used in the paper \cite{Denisenko et al 2008b} where our model of ionospheric conductivity
is described with stress on low latitudes.

Fig. \ref{fig:sig_120} demonstrates rather complicated global distribution at fixed moment of time.
The main reason for $\Sigma_{_P}$, $\Sigma_{_H}$ variations is the solar radiation.
We can see small values of $\Sigma_{_P}, \Sigma_{_H}$ (blue) in night time
which may be $2$ orders of magnitude less than their day time values.
As we see in Fig. \ref{fig:sig_120} the conductances $\Sigma_{_P}$, $\Sigma_{_H}$ are larger
in the Northern hemisphere. Northern polar cap is exposed to the solar radiation because we study a summer.
It is the model for $06$:$00$ UT in July. So the local midnight occurs around $\varphi_m=310^o$.
The second well seen singularity is the auroral enhancement.

The vicinity of the geomagnetic equator is also a specific domain.
The magnetic field lines are almost horizontal and so their long parts are embedded into the ionospheric layer
with large local conductivities $\sigma_{_P}$, $\sigma_{_H}$.

One can see an important difference between $\Sigma_{_P}$ and $\Sigma_{_H}$ near the geomagnetic equator.
Both of them increase but $\Sigma_{_P}$ decreases just at the equator as is shown
in Fig. \ref{fig:sig_120}.
The explanation can be found in Fig. \ref{fig:sigma} and \ref{fig:lines_ionosp} b.
Maximum of the local Hall conductivity $\sigma_{_H}$ is below $120\,\mbox{km}$ as is shown
in Fig. \ref{fig:sigma}.
So the last magnetic field line 4 in Fig. \ref{fig:lines_ionosp} b that is present
in the global pictures has $\Sigma_{_H}$ larger than $\Sigma_{_H}$
at the next lines (like the line 3) which are shown further from the equator.
Maximum of the local Pedersen conductivity $\sigma_{_P}$ is above $120\,\mbox{km}$.
So its integral along the line 4 is less than integral along some lines above it.
If we use the height $h=100\,\mbox{km}$ for presentation in Fig.
\ref{fig:sig_120} $\Sigma_{_H}$ also would have a minimum
at the equator and minimum of $\Sigma_{_P}$ would be much deeper.
If we chose height above $120\,\mbox{km}$ some magnetic field lines would disappear
and maximum of $\Sigma_{_H}$ would be lost.
So $h=120\,\mbox{km}$ looks optimal for presentation.
We already mentioned that it is a problem of visualization only and does not exist in calculations
because of special choice of the coordinate surface.

Pedersen conductivity $\Sigma_{_P}>0.1$ S in our model, it is
about $10$ S in middle latitude day-time ionosphere and increases up to $100$ S near the geomagnetic equator.
If a segment of the magnetic field line with nonzero
$\sigma_{_P}, \sigma_{_H}$ is short enough, then magnetic field lines are approximately parallel.
This is so in the high and middle latitude ionosphere where the magnetic field lines cross the ionospheric
conducting layer at some angle $\alpha$ from the horizon that increases the length of a segment of the line as
$\Delta l=\Delta h/\sin{\alpha}$, where $\Delta h$ is vertical size of the layer,
$\alpha$ is the magnetic inclination.
Our $\Sigma_{_P}, \Sigma_{_H}$ differ from height integrated parameters in high and middle latitudes
mainly by such a multiplier $1/\sin{\alpha}$.

\section{Magnetospheric conductor}
\label{sec:Magnetospheric conductor}

\begin{figure}[h]
\hspace*{15mm}
\includegraphics[scale=0.75]{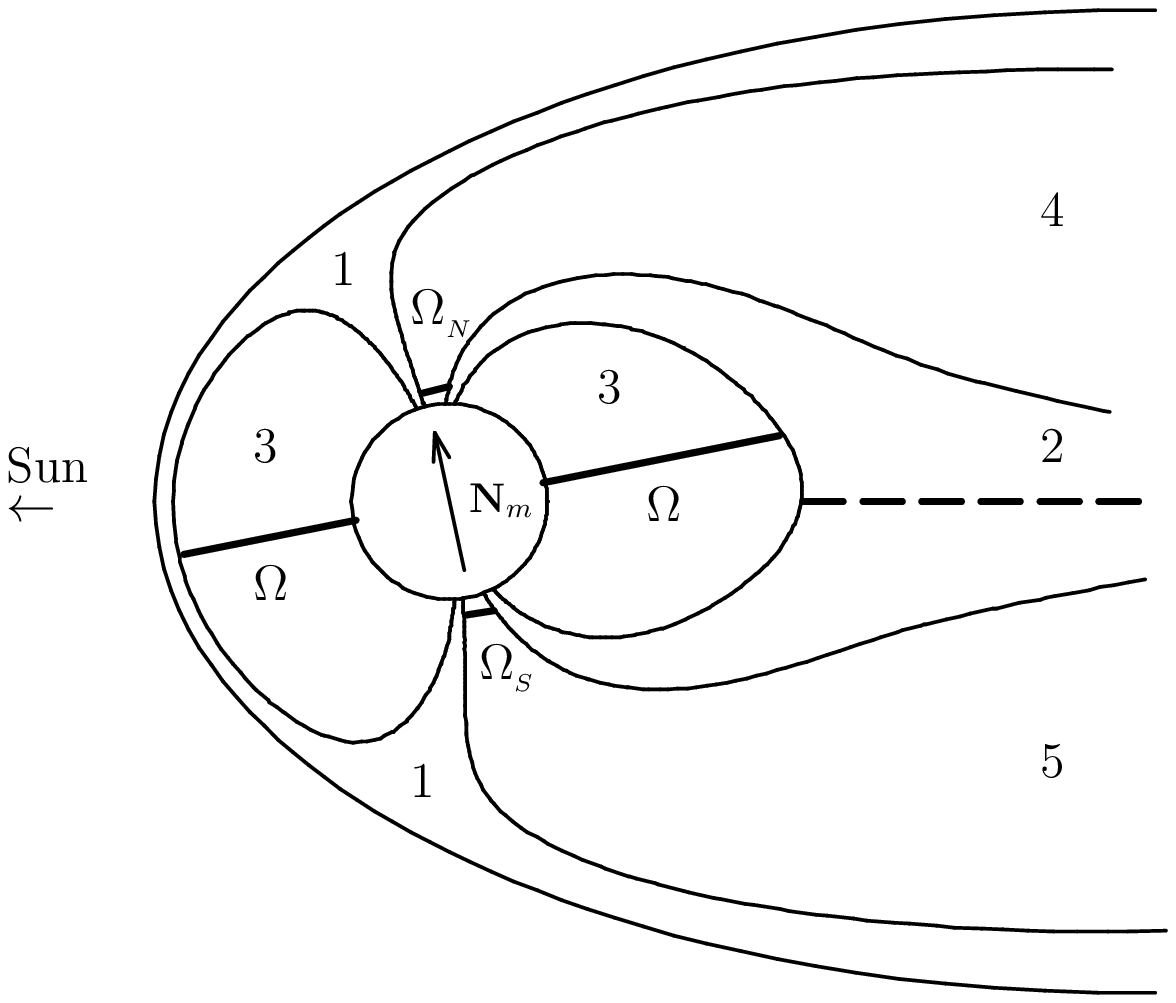}
\caption{
Diagram of the magnetosphere. Cross-section that contains the direction to the Northern magnetic pole ${\bf N}_m$
and direction to the Sun.
Magnetic field lines are plotted as the boundaries of 3-D domains which correspond to
vicinity of the magnetopause (1), the plasma sheet (2), closed magnetic field lines (3),
the Northern (4) and the Southern (5) polar cups.
Bold segments are the cross-sections of 2-D domains $\Omega_N$, $\Omega_S$, $\Omega$ which are used for calculations.
Dashed line - cross-section of the model of a thin plasma sheet.
}
\label{fig:magnetosphere}
\end{figure}

\begin{figure}[h]
\hspace*{45mm}
\includegraphics[scale=0.6]{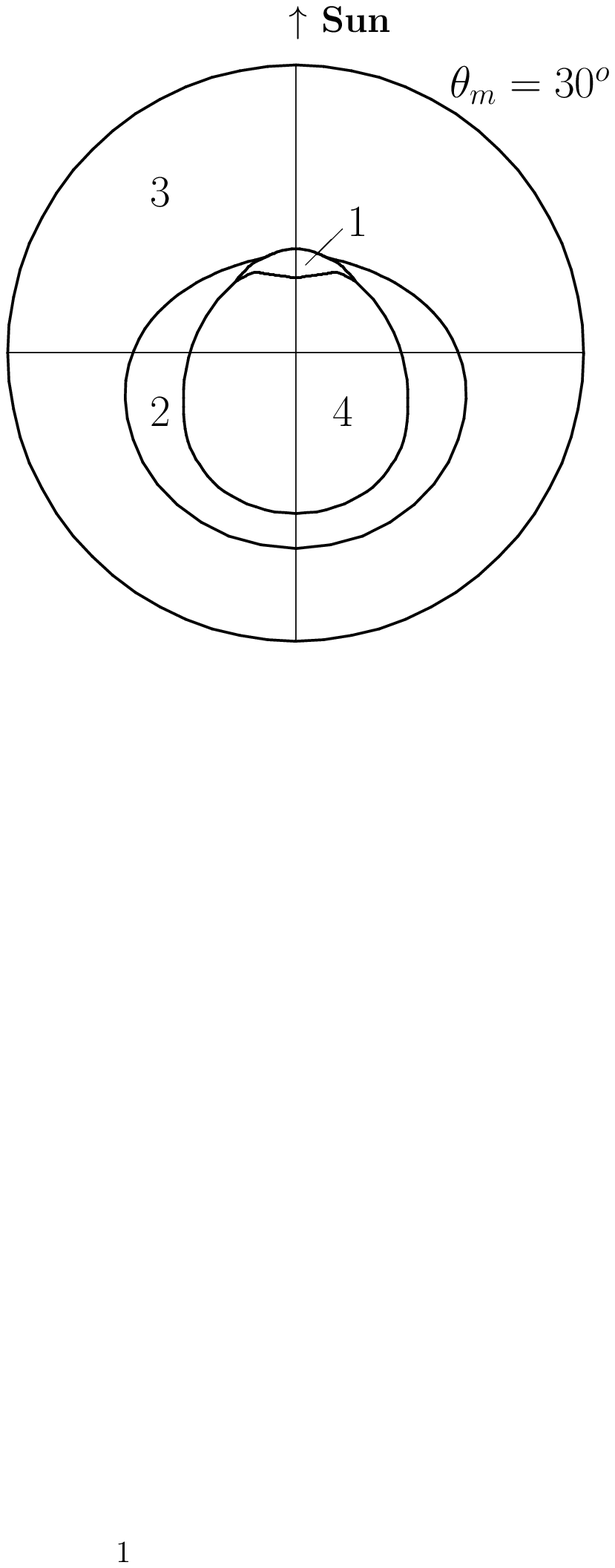}
\caption{
Mapping of the plasma sheet (2) and magnetopause (1) to the ionosphere.
(3) - closed magnetic field lines, (4) - polar cup with open magnetic field lines.
Reproduced from \cite{DenisenkoZamaiKitaev2003}.
}
\label{fig:cusp}
\end{figure}

\begin{figure}[h]
\hspace*{20mm}
\includegraphics[scale=0.8]{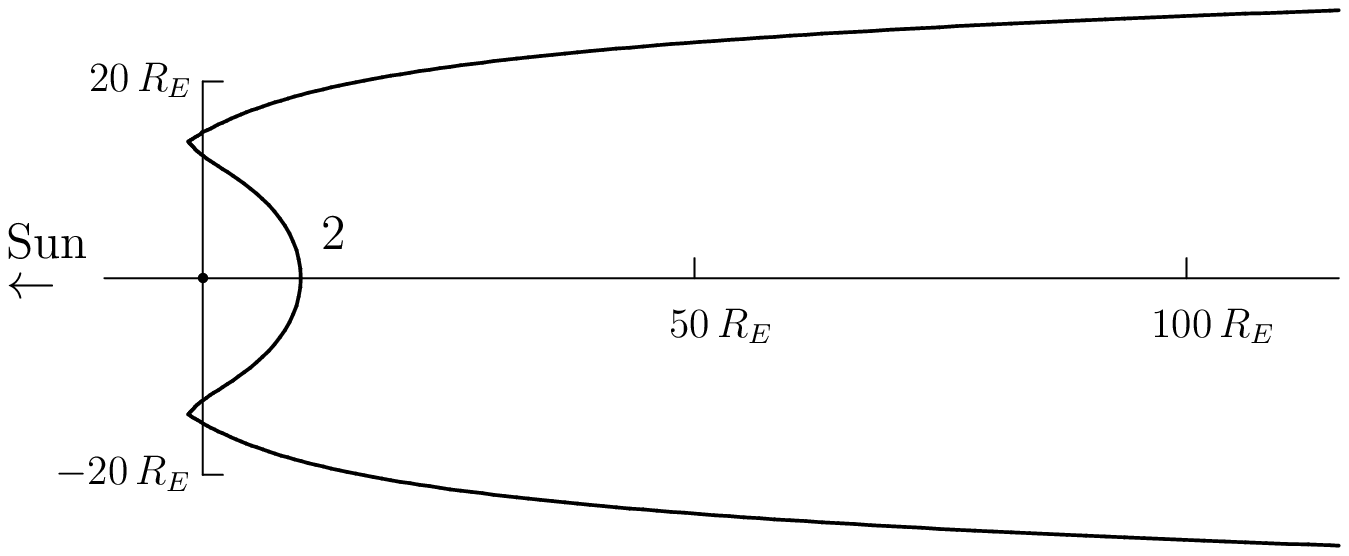}
\caption{
The model plasma sheet (2)
in the plane that is normal to the plane in Fig. \ref{fig:magnetosphere}
and contains the direction to the Sun (left horizontal).
The Earth is shown with the black dot at the coordinates origin.
Reproduced from \cite{DenisenkoZamaiKitaev2003}.
}
\label{fig:plsheet}
\end{figure}

A cross-section of the magnetosphere is schematically presented in Fig. \ref{fig:magnetosphere}.
The direction of the geomagnetic Northern pole is shown as ${\bf N}_m$ vector as is in summer.
The magnetosphere could be divided into 5 specific domains which are volumes of space.
Their cross-sections are shown in Fig. \ref{fig:magnetosphere}, \ref{fig:cusp}, \ref{fig:plsheet}.
The domain $\omega_1$ consists of magnetic field lines in vicinity of the magnetopause which correspond to cusp.
Its projection onto the ionosphere is shown as the domain (1) in Fig. \ref{fig:cusp}
reproduced from \cite{DenisenkoZamaiKitaev2003}.
The domain $\omega_2$ consists of magnetic field lines which correspond to the plasma sheet.
Its projection onto the ionosphere is shown as the domain (2) in Fig. \ref{fig:cusp}.
The domain $\omega_3$ consists of closed magnetic field lines.
Its project to the ionosphere is shown as the domain (3) in Fig. \ref{fig:cusp}.
The domains $\omega_4$ and $\omega_5$ consist of magnetic field lines which go to the distant tail
from Northern and Southern polar caps respectively.
The Northern polar cap is shown as the domain (4) in Fig. \ref{fig:cusp}.

As is seen in Fig. \ref{fig:sigma} the conductivities
$\sigma_{_P}, \sigma_{_H}$ are small above $500$ km.
This is so for the whole magnetosphere with exclusion of the plasma sheet
in the magnetospheric tail and vicinity of the magnetopause.
Space distribution of conductivity in the plasma sheet is not known.
Fortunately estimations \cite{Cattell 1996} are enough.
Fig. \ref{fig:plsheet} presents the model \cite{DenisenkoZamaiKitaev2003}
that simulates the plasma sheet as a thin layer.
The cross-section of this thin layer is shown with dashed line in Fig. \ref{fig:magnetosphere}.
In accordance with \cite{Cattell 1996} the integral of conductivity across the layer that means
normal to the plane in Fig. \ref{fig:plsheet} is of about $100\,\mbox{S}$.
Roughly speaking the plasma sheet in Fig. \ref{fig:plsheet}
has shape of a quadrangle with width $40\,R_{_E}$ and length more than $200\,R_{_E}$
in direction away from the Sun. It means that current across the plasma sheet (normal to direction Earth - Sun)
would be of about $500\,\mbox{A}$ because of voltage $1\,\mbox{V}$.

Since conductivity along magnetic field lines is large, such a voltage is the same
as in the ionospheric projection of the plasma sheet
shown as the domain (2) in Fig. \ref{fig:cusp}.
If there is $1\,\mbox{V}$ voltage along the strip (2) in the ionosphere there would be $500\,\mbox{A}$
current from one its end to another
through the plasma sheet. The same electric current because of the same voltage $1\,\mbox{V}$ would
flow by this strip in the ionosphere if Pedersen conductance is about $7500\,\mbox{S}$ in this ionospheric strip
since its length is about $15$ time larger than width.

Similar conclusions can be made about the magnetopause. More or less conventional estimation
of the magnetic Reynolds number is of about $10^4$ near the magnetopause \cite{DenisenkoZamaiKitaev2003}
that means conductivity $2\cdot 10^{-4}\,S/m$.

Within $500\,\mbox{km}$ thickness of the magnetopause it gives about $100\,\mbox{S}$ conductance
in tangential to the magnetopause directions.
It also is equivalent to large conductivity in the cusp that is shown as the domain (1) in Fig. \ref{fig:cusp}.

Of course these estimations are approximate.
Nevertheless they correspond to the usual approach in physics of magnetosphere in accordance to which
the distribution of potential in polar parts of the auroral zones is defined by magnetospheric generators.
Since usually there is morning - evening electric field in the polar caps the main parameter
of such a potential distribution
is so called polar-cap potential drop that can reach $100\,\mbox{kV}$ during magnetospheric substorms
\cite{Hargreaves}.
We already mentioned that electric fields and currents of these magnetospheric generators are not present in our model.
Anyway it means that the interior resistances of these generators are much smaller than the resistances of the ionospheric regions
which are connected with them by magnetic field lines.
For ionospheric generators it means that corresponding magnetospheric objects are good conductors and
they can be approximately regarded as ideal conductors.

We idealize such a huge conductance as infinity that means constant potential value in both domains
$\omega_1 \&\,\omega_2$. We can define this constant as zero:
\begin{equation}
\left.V\right|_{\omega_1 \&\,\omega_2}=0.
\label{eq:7.1}
\end{equation}

This condition cuts the ionosphere into three parts which are the Northern and the Southern polar caps and the main part
that is the project of the 3-D domain $\omega_3$ of closed magnetic field lines.
These parts of the ionospheric conductor in many aspects can be analyzed independently
since potential at their boundaries is defined as zero beforehand as consequence of (\ref{eq:7.1}).
We use (\ref{eq:7.1}) to set boundary conditions in the next section 6.

While integral conductances are calculated as is described in previous section,
we use IGRF model of geomagnetic field.
This model presents only the field that is created by currents which exist inside the Earth.
Such a field is dominant in the ionosphere, but it decreases
in the magnetosphere, and tracing of the magnetic field lines needs addition of the fields
produced by magnetospheric currents.
First of all they are currents at the magnetopause which close the geomagnetic field inside
the magnetosphere and
currents in the current sheet which pull magnetic field into the tail.
It seems to us that the best empirical model of
magnetospheric magnetic field is created by Tsyganenko \cite{Tsyganenko}.
Since we are interested only in closed magnetic field lines
(region 3 in Fig. \ref{fig:magnetosphere}) which are not extended too far from the Earth,
we use our more simple model \cite{Radio_Emissions_2006}
in addition to the field of the model IGRF.
Tracing of magnetic field lines is necessary to find conjugated points
in the Northern and Southern hemispheres since
these points have equal potential because of high conductivity along a magnetic field line.
It means that these conductors in the Northern and Southern hemispheres are connected in parallel and one must
take this circumstance into account while ionospheric electric field and current simulation
as we describe in next section.

We also use a simple approximation for the boundary of the region 3 in Fig. \ref{fig:cusp} with closed magnetic field lines.
The regions 2, 3 together are represented as a ring with width $\delta\theta_m=5^o$ that occupies central half
of the auroral zone shown in Fig. \ref{fig:sig_120_caps}.
The same boundary in the Southern hemisphere is obtained by mapping along magnetic field lines.
The boundary of the Northern polar caps, that is the domain 4 in Fig. \ref{fig:magnetosphere}, \ref{fig:cusp}, is defined
as the polar boundary of this ring. It corresponds to the shift of $5^o$ in direction to the pole from the boundary of the region 3.
The same shift is done in the Southern polar cap.
Strictly speaking positions of these boundaries are not well definite since they are defined on the base of a rough model of the plasma sheet.
So we must analyze their influence on the results. In some important cases it is small.

\section{The charge conservation law for 2-D model of the ionospheric conductor} 
\label{sec:charge_cons_2-D}

The ionospheric layer $h_{_I}<h<h_{_M}$, in which the conductivity across the magnetic field is concentrated,
is sufficiently thin, so the neighbor magnetic field lines are assumed to be parallel,
as shown in Fig. \ref{fig:lines_ionosp} a.
The angle between ${\bf B}$ and the horizon is designated as $\alpha$, it is the magnetic inclination.
We introduce local Cartesian coordinates with the axis $z'$ along ${\bf B}$ and denote the coordinates
of the bottom and top points through $z'_{_I}$ and $z'_{_M}$.
The horizontal normal to ${\bf B}$ is used as the $y'-$axis.
Then the $x'-$axis lies in the plane of Fig. \ref{fig:lines_ionosp} a.

Then the charge conservation law (\ref{eq:3}) can be written as
\begin{equation}
\frac{\partial j_{x'}}{\partial x'}+\frac{\partial j_{y'}}{\partial y'}+\frac{\partial j_{z'}}{\partial z'}=0,
\label{eq:13.1}
\end{equation}
where the right-hand side is zero, since we do not consider ionospheric external currents.
We integrate this equation in $z'$ from $z'_{_I}$ to $z'_{_M}$ and express the value of the last integral:
\begin{equation}
\frac{\partial}{\partial x'}\int j_{x'}dz'+\frac{\partial}{\partial y'}\int j_{y'}dz'+j_{z'}(x',y',z'_{_M})-j_{z'}(x',y',z'_{_I})=0,
\label{eq:13.2}
\end{equation}

The first two integrals are the components of the vector ${\bf J}_{_\perp}$ (\ref{eq:20.1})
the source of which are thunderstorms in our model.
The last term describes the current from the atmosphere to the ionosphere.
This vertical current density $j_{ext}(\lambda_m,\varphi_m)$ provides the same charge flow to the ionosphere as the current with density
\begin{equation}
j_{x'}=j_{y'}=0, \quad j_{z'}=j_{ext}/\sin{\alpha}.
\label{eq:13.3}
\end{equation}

A similar penultimate term in (\ref{eq:13.2}) describes the current from the ionosphere to the magnetosphere.
We denote the equivalent density of the vertical current by $j_{_M}$.
Since we do not consider magnetospheric generators, it is zero in polar caps,
from which the magnetic field lines go to infinity through the domains $\omega_4$ and $\omega_5$ in which there is no conductivity
across ${\bf B}$.
In the main part of the ionosphere, this term describes the exchange of charges between conjugate points.

Summarizing the above, equation (\ref{eq:13.2}) can be written in the form
\begin{equation}
\frac{\partial}{\partial x'}J_{x'}+\frac{\partial}{\partial y'}J_{y'}=(j_{ext}-j_{_M})/\sin{\alpha}.
\label{eq:13.4}
\end{equation}

By virtue of Ohm's law (\ref{eq:27}-\ref{eq:29})
\begin {equation}
\left(\begin{array}{c}
J_{x'} \\
J_{y'}
\end{array}\right)=
\left(\begin{array}{cc}
\Sigma_{_P} & -\Sigma_{_H} \\
\Sigma_{_H} &  \Sigma_{_P}
\end{array}\right)
\left(\begin{array}{c}
E_{x'} \\
E_{y'}
\end{array}\right),
\label{eq:13.5}
\end {equation}
where the components of the electric field strength can be expressed in terms of the potential according to the formula (\ref{eq:16})
\begin{equation}
E_{x'}=-\frac{\partial V(x',y')}{\partial x'}, \quad E_{y'}=-\frac{\partial V(x',y')}{\partial y'}.
\label{eq:13.6}
\end{equation}

This record takes into account the constancy of the potential on the entire magnetic field line,
that is, its independence from $z'$.
Taking into account (\ref{eq:13.5}, \ref{eq:13.6}) the equation (\ref{eq:13.4}) takes the form
\begin{equation}
-\frac{\partial}{\partial x'}\left(\Sigma_{_P}\frac{\partial V}{\partial x'}
-\Sigma_{_H}\frac{\partial V}{\partial y'}\right)
-\frac{\partial}{\partial y'}\left(\Sigma_{_H}\frac{\partial V}{\partial x'}
+\Sigma_{_P}\frac{\partial V}{\partial y'}\right)
=\frac{j_{ext}-j_{_M}}{\sin{\alpha}}.
\label{eq:13.0}
\end{equation}

With the help of simple but cumbersome formulae, one can go from the local coordinates $x', y'$
to the magnetospheric coordinates $x_m, y_m$ of the same magnetic field line
on the plane with a fixed value of $z_m$. The details of this geometric transformation
are given in \cite{Denisenko et al 2008b}.
Since magnetic field lines for non dipolar magnetic field can be traced only numerically
this transformation is also done numerically.
Then the equation (\ref{eq:13.0}) takes the form
\begin{equation}
-\frac{\partial}{\partial x_m}\left(\Sigma_{xx}\frac{\partial V}{\partial x_m}
+\Sigma_{xy}\frac{\partial V}{\partial y_m}\right)
-\frac{\partial}{\partial y_m}\left(\Sigma_{yx}\frac{\partial V}{\partial x_m}
+\Sigma_{yy}\frac{\partial V}{\partial y_m}\right)
=Q_{ext}-Q_{_M}.
\label{eq:13.7}
\end{equation}

The potential $V(x_m,y_m)$ is an unknown function of two variables, and the coefficients of the conductivity
tensor $\hat\Sigma$ and the function $Q_{ext}$ are given.

For the Northern and Southern polar 3-D domains $\omega_4$ and $\omega_5$
this equation must be satisfied in all points
of the flat 2-D domains $\Omega_N$ and $\Omega_S$ correspondingly.
Their cross-sections are shown in Fig. \ref{fig:magnetosphere}.
As already noted, the second term on the right-hand side is zero when considering polar caps.

In the main part of the ionosphere above $90$ km height the entire magnetic field line has the same potential.
We can write down charge conservation law for the entire magnetic field line
by summing equations of the form (\ref{eq:13.7}) obtained for the halves of this line.
Since the function $V(x_m,y_m)$ is the same in them the equation keeps its shape with summing of
the coefficients of the conductivity tensor $\hat\Sigma$ from two hemispheres.
The last term $Q_{_M}$ has the same value and opposite sign in both hemispheres
because it represents the same current along the magnetic field line at its opposite ends.
For this 3-D domain $\omega_3$ this equation must be satisfied in all points
of the flat 2-D domain $\Omega$ which cross-section is shown in Fig. \ref{fig:magnetosphere}.
Thus the equation (\ref{eq:13.7}) has the same shape in all three flat domains $\Omega_N$, $\Omega_S$ and $\Omega$
\begin{equation}
-\frac{\partial}{\partial x_m}\left(\Sigma_{xx}\frac{\partial V}{\partial x_m}+\Sigma_{xy}\frac{\partial V}{\partial y_m}\right)
-\frac{\partial}{\partial y_m}\left(\Sigma_{yx}\frac{\partial V}{\partial x_m}+\Sigma_{yy}\frac{\partial V}{\partial y_m}\right)
=Q_{ext}.
\label{eq:13.8}
\end{equation}

Symmetrical part of the conductivity tensor $\hat\Sigma$
in (\ref{eq:13.5}) is a positive definite one since $0<\Sigma_{_P}<\infty$ in the domains of interest.
For a physicist it means the natural property of positiveness of the Joule dissipation.
From mathematical point of view it is an important property of the coefficients
of the partial differential equation (\ref{eq:13.5}).
These property is not corrupted by the used coordinate transformation and symmetrical part of the tensor
$\hat\Sigma$ in (\ref{eq:13.8}) also is a positive definite one.
So the partial differential equation (\ref{eq:13.8}) has an elliptical type
which means that one can use boundary conditions similar to ones for Poisson equation.

\section{Boundary value problems} 
\label{sec:Boundary_value_problems}

Let us start with Northern polar cap. We use the plane $z_m=z_{_N}$ with some value $z_{_N}$ greater than Earth's radius.
The 3-D domain $\omega_4$ crosses this plane in 2-D domain $\Omega_N$
which cross-section is plotted with bold line in Fig. \ref{fig:magnetosphere}.
It would be a circle if the geomagnetic field were axially symmetrical.
We denote its boundary line as $\Gamma_N$ where the potential equals zero because of (\ref{eq:7.1})
\begin{equation}
\left.V\right|_{\Gamma_N}=0.
\label{eq:8.0}
\end{equation}

So we are to solve Dirichlet boundary value problem (\ref{eq:13.8}, \ref{eq:8.0})
for unknown function $V(x_m,y_m)$ in 2-D flat domain $\Omega_N$.
Such a problem has unique solution since the partial differential equation (\ref{eq:13.8}) is an elliptical one.
We solve similar boundary value problem for the Southern polar cap:
\begin{equation}
\left.V\right|_{\Gamma_S}=0.
\label{eq:8.1}
\end{equation}

For the main part of the ionosphere the 2-D domain $\Omega$ is in the plane $z_m=0$,
which cross-section is also shown in Fig. \ref{fig:magnetosphere}.
For dipolar magnetic field $\Omega$ would be a axially symmetrical ring.
Its outer boundary $\Gamma_{aur}$ corresponds to the boundary magnetic field lines which neighbor lines
are in 3-D domains $\omega_1$ or $\omega_2$ in which potential equals zero (\ref{eq:7.1}).
Therefore the same boundary condition as (\ref{eq:8.0}) can be used at $\Gamma_{aur}$
\begin{equation}
\left.V\right|_{\Gamma_{aur}}=0.
\label{eq:8.6}
\end{equation}

The interior boundary $\Gamma_{eq}$ of $\Omega$ corresponds to the last magnetic field lines which are regarded
as ionospheric ones and so as equipotential ones.
For simplicity and clarity, we first consider the points of this boundary near which the magnetic field
has the form shown in Fig. \ref{fig:lines_ionosp} b.
A dark segment is a cross-section of a part of the domain $\Omega_N$ that is near the boundary $\Gamma_{eq}$.
The lower point of this segment belongs to the considered boundary.
It means that the magnetic field line $5$ is the last one.
Above it there is the ionosphere, in which the approximation $\sigma_{_\parallel}=\infty$ (\ref{eq:14}) is used,
which made it possible to construct a 2-D model and formulate the equation (\ref{eq:13.8}).

At the height $h_{_I}$, we must know a global distribution of the current density
from the atmosphere to the ionosphere $j_{ext}(\lambda_m,\varphi_m)$.
Let us shift these values in a small area from the horizontal surface $h=h_{_I}$ to the curve surface which consists of the lines like line 5.

If we select the last magnetic field line $5$ below actual position, the integral conductivity $\Sigma_{_P}$
on it will be smaller.
If the top of this line drops to a height of $h_{_I}$, and hence the length of its segment above $h_{_I}$
becomes zero,
$\Sigma_{_P}=0$, since $\Sigma_{_P}$ is obtained by integrating $\sigma_{_P}$ over this segment.
The vanishing of the coefficient $\Sigma_{_P}$ in equation (\ref{eq:13.8}) radically changes
the type of the equation.
Decrease of $\Sigma_{_P}$ and its approach to zero degrades the properties of the equation,
which are used both in justifying the correctness of the statement of the boundary value problem,
and in its numerical solution.

When the lowest magnetic field line is selected, the charge conservation law can be used.
The current into it from below equals to the integral over this segment of the density
of the atmospheric current $j_{ext}(\lambda_m,\varphi_m)$.
We denote it by $J^0_{eq}$.
The current flowing into this line from above, that is, from within $\Omega_N$, is determined
by Ohm's law (\ref{eq:13.5}):
\begin {equation}
J_{y'}=\Sigma_{_H}E_{x'}+\Sigma_{_P}E_{y'}.
\label{eq:8.2}
\end {equation}

To fulfill the charge conservation law, the sum of these currents must be zero.
This condition can be converted to the coordinates $x_m,y_m$, as was done in the transition
from the equation (\ref{eq:13.0}) to (\ref{eq:13.8}). We get:
\begin{equation}
\left.J_{\nu}\right|_{\Gamma_{eq}}=-J^0_{eq},
\label{eq:8.7}
\end{equation}
where the subscript $\nu$ denotes the current component normal to the boundary.
It is vertical in our case.
In view of (\ref{eq:8.2}) and similar expression for $J_{\nu}$ in coordinates $x_m,y_m$ this boundary condition
defines the value of the inclined derivative of $V(x_m,y_m)$.

Thus we obtain the separate boundary value problem of mixed type (\ref{eq:13.8}, \ref{eq:8.6}, \ref{eq:8.7})
in the flat ring $\Omega$ with boundaries $\Gamma_{aur}$, $\Gamma_{eq}$.
It represents the main part of the ionosphere.
Such a problem has unique solution \cite{book}.

Strictly speaking, it is necessary to solve the 3-D problem of electrical conductivity
in the considered lower part of the equatorial ionosphere.
However, if one is not interested in the detailed distributions of electric fields and currents inside it,
it is sufficient to take it into account approximately, for example in the proposed way.
The point is that this region only slightly changes the total conductivity of the region encompassing
lines 1-5 in Fig. \ref{fig:lines_ionosp} b, that is the region of equatorial electrojets.
In other words, the equatorial electrojets would be only slightly changed if the conductivity is changed below
the level chosen sufficiently low.
We make this choice based on test calculations.
Moreover it is possible to avoid the detailed simulation of the entire region covering the lines 1-5,
setting a special boundary condition on the line 1 in Fig. \ref{fig:lines_ionosp} b \cite{book}.

\section{Conclusions} 
\label{sec:Conclusions}

A model of the ionospheric global conductor is described.
The ionospheric conductor is considered in the framework of a two-dimensional approximation based
on high conductivity in the direction of the magnetic field.
Under this assumption the magnetic field lines are equipotential,
and charge transfer between them is determined only by integral Pedersen and Hall conductivities.
The model is the first approximation in the small parameter expansion of the solution of three-dimensional problems of electrical conductivity.
The small parameter is the ratio of Pedersen and field-aligned conductivities.
The space distributions of the Pedersen and Hall conductivities are calculated using the empirical models IRI, MSISE, IGRF
and applied to construct the maps of the integral conductivities.
This method of reducing the 3-D model to a 2-D model has clear advantages \cite{SPb_2016_ion} in
comparison with the use of a small thickness of the ionospheric layer, in which the conductivity is concentrated across
magnetic field \cite{Hargreaves}.

Such a 2-D model was used in \cite{Ampferer et al 2010} to calculate the local electric fields in the ionosphere
penetrating from the atmosphere.
It is shown in \cite{Denisenko et al 2013} that the reducing the 3-D to a 2-D model adds only a small error
when the horizontal scale of the process exceeds $100$ km.

The parts of the magnetosphere with high conductivity across the magnetic field lines,
namely, the cusps and the plasma layer are analyzed.
It is shown that the connection of these magnetospheric conductors
in parallel to the ionosphere makes the auroral zones equipotential regions.
As a consequence, for the ionospheric electric fields, which generators
are located in the ionosphere or in the atmosphere, the global problem of electrical conductivity is separated into three
independent boundary value problems in three regions: two polar caps and the main part of the ionosphere which includes
the mid- and low-latitude parts of the ionosphere.

The model can be used for the analysis of the ionospheric part of the Global Electric Circuit.
Also the model is applicable for calculation of the ionospheric dynamo electric field.
In general, the created model can be used in more complex ionospheric models.
It also allows to take into account the ionospheric conductor in the models of the magnetosphere.
In the latter case, it allows to find the ionospheric potential distribution for a given distribution
of the field-aligned currents flowing into the ionosphere from the magnetosphere.
In this case, the auroral zones and the polar caps
are included to the global boundary value problem of the electrical conductivity of the ionosphere.
This kind of models of the field-aligned currents can be constructed
by empirical models of the field-aligned currents as we did in \cite{planet}.
Taking into account the connection between the conjugate parts of the mid-latitude ionosphere and
the low-latitude ionosphere leads to the problem in a non-schlicht domain \cite{non-schlicht}.

A numerical method is proposed for the created model in \cite{book,Multigrid}.

The model is simplified if we suppose the geomagnetic field to be dipole.
In that approximation, a number of problems on the generation of ionospheric electric fields by magnetospheric generators
have been solved \cite{planet,DenisenkoZamaiKitaev2003}.

\section*{Acknowledgments}

The research is supported by Russian Foundation for Basic Research (project 18-05-00195).
The author is grateful to Prof. Michael Rycroft for the fruitful discussions.

\end{document}